\documentclass[12pt,amsmath,amssymb]{revtex4}
\usepackage{graphics}
\usepackage{amsfonts}
\usepackage{amsmath}
\usepackage{epsfig}
\usepackage{epsf}
\usepackage{graphicx}
\usepackage{dcolumn}
\usepackage{bm}
\usepackage{color}
\usepackage{array}
\usepackage{amssymb}

\newcommand {\sla}[1]{ #1 \!\!\!/}

\begin{document}

\title{Two-photon exchange effects in $e^+e^- \rightarrow \pi^+\pi^-$ and time-like pion electromagnetic form factor }
\author{
Hong-Yu Chen, Hai-Qing Zhou \protect\footnotemark[1] \protect\footnotetext[1]{E-mail: zhouhq@seu.edu.cn} \\
School of Physics,
Southeast University, NanJing 211189, China}

\date{\today}

\begin{abstract}
The two-photon-exchange (TPE) effects in the process $e^+e^- \rightarrow \pi^+\pi^-$ at large momentum transfer are discussed  within the perturbative QCD (pQCD). The contributions from the twist-2 and twist-3 distribution amplitudes (DAs) of pion are considered in the estimation. Different with the results under the one-photon-exchange (OPE) approximation, the TPE effects result in an asymmetry of the differential cross section on the scattering angle. The precise measurement of this asymmetry by the further experiment is an precise test of pQCD at large momentum transfer.  The time-like electromagnetic form factor of pion at the leading order of pQCD is re-discussed and the comparison of our results with those in the references are presented.
\end{abstract}

\maketitle

\section{Introduction}
The pion and the proton are the most elemental bound states due to the strong interaction.
The knowledge on their structures is important to test our understanding on QCD. The electromagnetic (EM) form factor
is one of the most simple and naive non-perturbative quantity reflecting the structures of these bound states.

In 2000, the measurements of ratio of the EM form factors of proton by the polarized method \cite{Jones2000,Gayou2002} give very different results with those given by Rosenbluth method \cite{Andivahis1994,Walker1994}.  This suggests the extraction of the EM form factors from the experimental data is a non-trivial problem. The two-photon-exchange (TPE) effects in the unpolarized $ep$ scattering are expected to explain the discrepancy between the results by the polarized method and Rosenbluth method.  Many theoretical methods have been used to estimate the TPE effects such as  the hadronic model
\cite{Blunden03,Kondra05,Blunden05,zhouhq2014}, GPD method \cite{Chen04,Afana05}, phenomenological parametrizations \cite{Chen07,BK07}, dispersion relation approach \cite{BK06,BK08,BK11,BK12,BK14,Blunden2017}, pQCD calculations \cite{BK09,Kivel09} and SCEF method \cite{TPE-SCEF}. The recent experimental results on the $R^{2\gamma}\equiv \sigma_{e^{+}p\rightarrow e^{+}p}/\sigma_{e^{+}p\rightarrow e^{-}p}$ \cite{OLYMPUS2017} which measures the TPE effect directly shows the estimation by the most recent calculation \cite{Blunden2017} does not match the experimental data very well.  All these mean our understanding on the TPE effects in the $ep$ scattering still needs to be imporved both in the theoretical and the experimental aspects.

The TPE effects in the other processes also abstract many interesting and are discussed in references, for example $e^{+}e^{-} \rightarrow p \overline{p}$ \cite{DianYongChen2008}, $e\pi$ scattering \cite{Blunden2010,YuBingDong2010} and unpolarized $\mu p$ scattering \cite{DianYongChen2013,Tomalak2014,Afanasev2016,zhouhaiqing2017}.  In literatures, the TPE effects in the process $e^{+}e^{-} \rightarrow \pi^{+}\pi^{-}$ is usually ignored since the TPE effects will not affect the total cross section or the time-like EM form factor of pion, while the TPE effects still play their role in the angle dependent of the differential cross section. The EM form factor of  pion in the space-like region at high momentum transfer has played important role in the test of  pQCD factorization \cite{ZhengTaoWei2003,XingGangWu2004,Raha2009-PRD}, while the experimental measurement of the EM form factor of pion in the space-like region is not a trivial problem since there is no pion target.  The study of the EM form factor of pion in the time-like region is another window to test the pQCD factorization \cite{Raha2010-PLB,HaoChungHu2013-PLB,ShangChen2015-PLB}. The study of the TPE effects in this process also play the similar role to test the pQCD factorization and to help us understand the TPE effects. In this work, we estimate this effect and we also clarify some discussion on the time-like EM form factor of pion at the leading order of pQCD given in literatures. We arrange our work as following, in Section II we give a simple introduction on the cross section of $e^{+}e^{-} \rightarrow \pi^{+}\pi^{-}$ and the time-like EM form factor of pion by pQCD under the one-photon-exchange (OPE) approximation, in Section III we discuss the TPE effects in this process, in Section IV we discuss the input used in our practical estimation, and in Section V we give the numerical results and our conclusion.

%
%
%

\section{$e^+e^-\rightarrow \pi^+\pi^-$ via one-photon-exchange}
\begin{figure}[htbp]
\center{\epsfxsize 3.4 truein\epsfbox{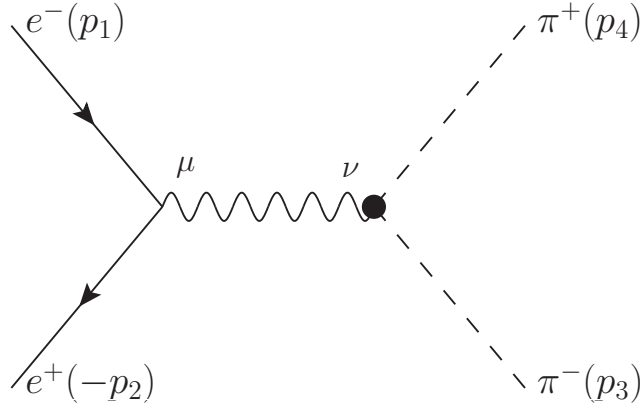}}
\caption{Diagrams for $e^{+}e^{-}\rightarrow \pi^{+}\pi^{-}$ with one-photon exchange (OPE).}
\label{figure:Amp-OPE}
\end{figure}
In the OPE approximation, the process $e^+e^- \rightarrow \pi^+\pi^-$  can be described by the diagram showed in Fig.\ref{figure:Amp-OPE} and the corresponding amplitude can be expressed as
\begin{eqnarray}
\mathcal{M}^{1\gamma}=[{\bar u}(-p_2,m_e)(-ie\gamma^\mu) u(p_1,m_e)]D_{\mu\nu}(q)[-ie (p_4-p_3)^\nu F_{\pi}(s)],
\label{Amp1}
\end{eqnarray}
where $p_1,p_2,p_3$ and $p_4$ are the momenta of the initial electron, initial anti-electron, finial $\pi^{-}$ and $\pi^{+}$, $D_{\mu\nu}(q)$ is the photon propagator, $q=p_1+p_2=p_3+p_4$, $Q^2=q^2$ and $F_{\pi}(Q^2)$ is the time-like EM form factor of pion which is defined as
\begin{eqnarray}
<\pi^+\pi^-|j_{\mu}(0)|0>\equiv -(p_4-p_3)_{\mu} F_{\pi}(Q^2),
\label{definition-FF}
\end{eqnarray}
with $j_{\mu}=\sum e_i\overline{q}_i\gamma_{\mu}q_i$, $q_i$ the quark fields, $i$ the flavor indexes of the quarks and $e_i$ the corresponding electric charge ($-1$ for electron).

By Eq.(\ref{Amp1}), the unpolarized differential cross section can be expressed as
\begin{eqnarray}
d\sigma_{un}^{1\gamma}=\frac{1}{2}e^2F_\pi(Q^2)\ F^*_\pi(Q^2)\ sin^2\theta,
\end{eqnarray}
where $\theta$ is the angle between the three momenta of initial electron(${\bf p}_1$) and finial $\pi^{-}({\bf p}_3)$ in the center frame, $e=-|e|=-\sqrt{4\pi\alpha_{QED }}$.

\begin{figure}[htbp]
\center{\epsfxsize 2.5 truein\epsfbox{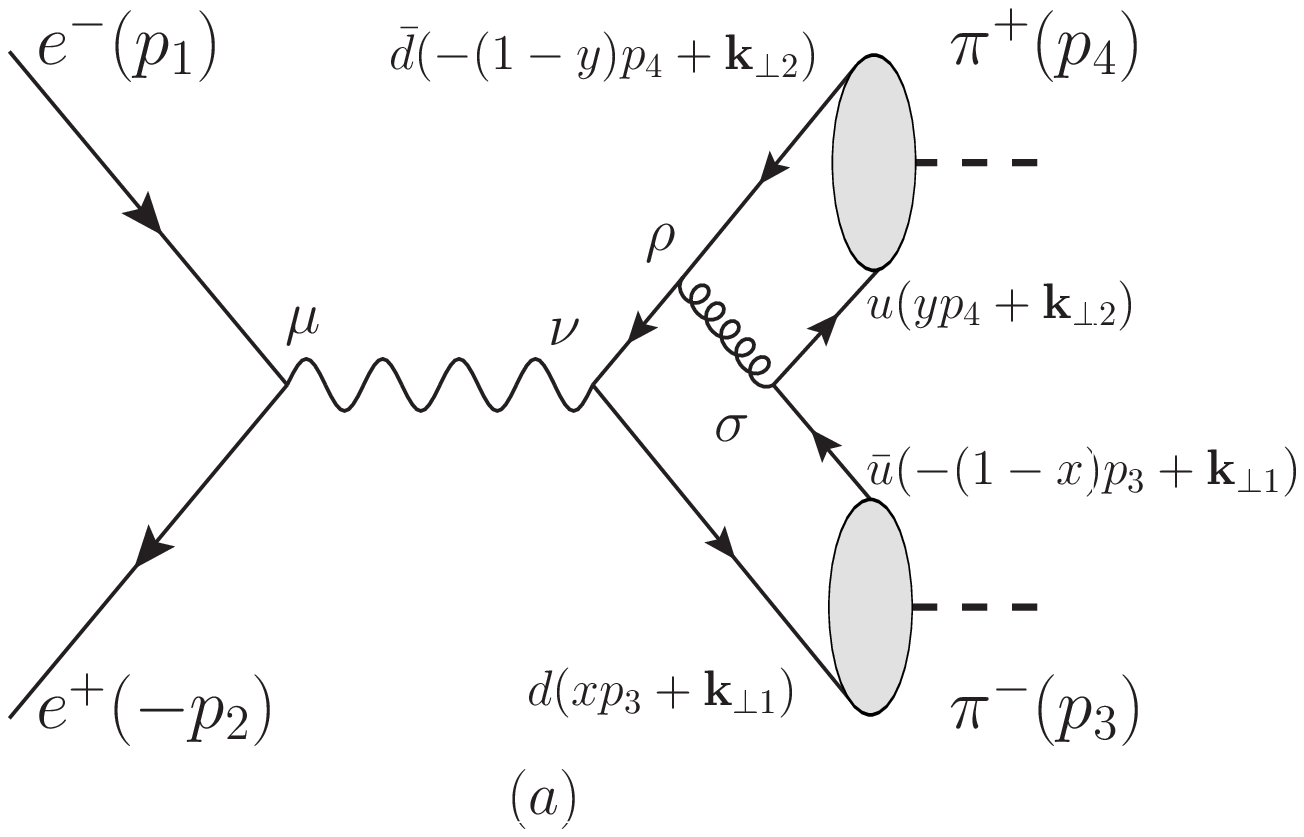}\epsfxsize 2.8 truein\epsfbox{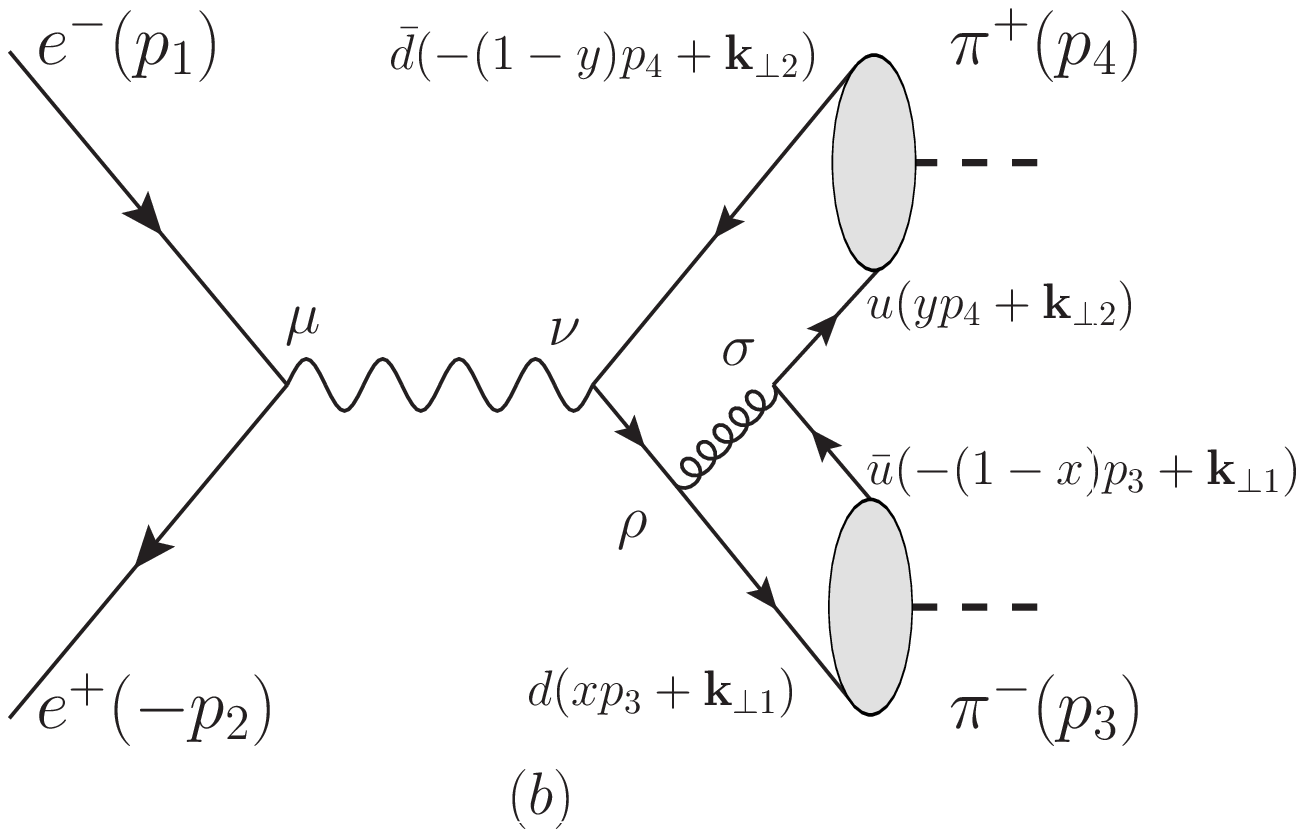}}
\center{\epsfxsize 2.5 truein\epsfbox{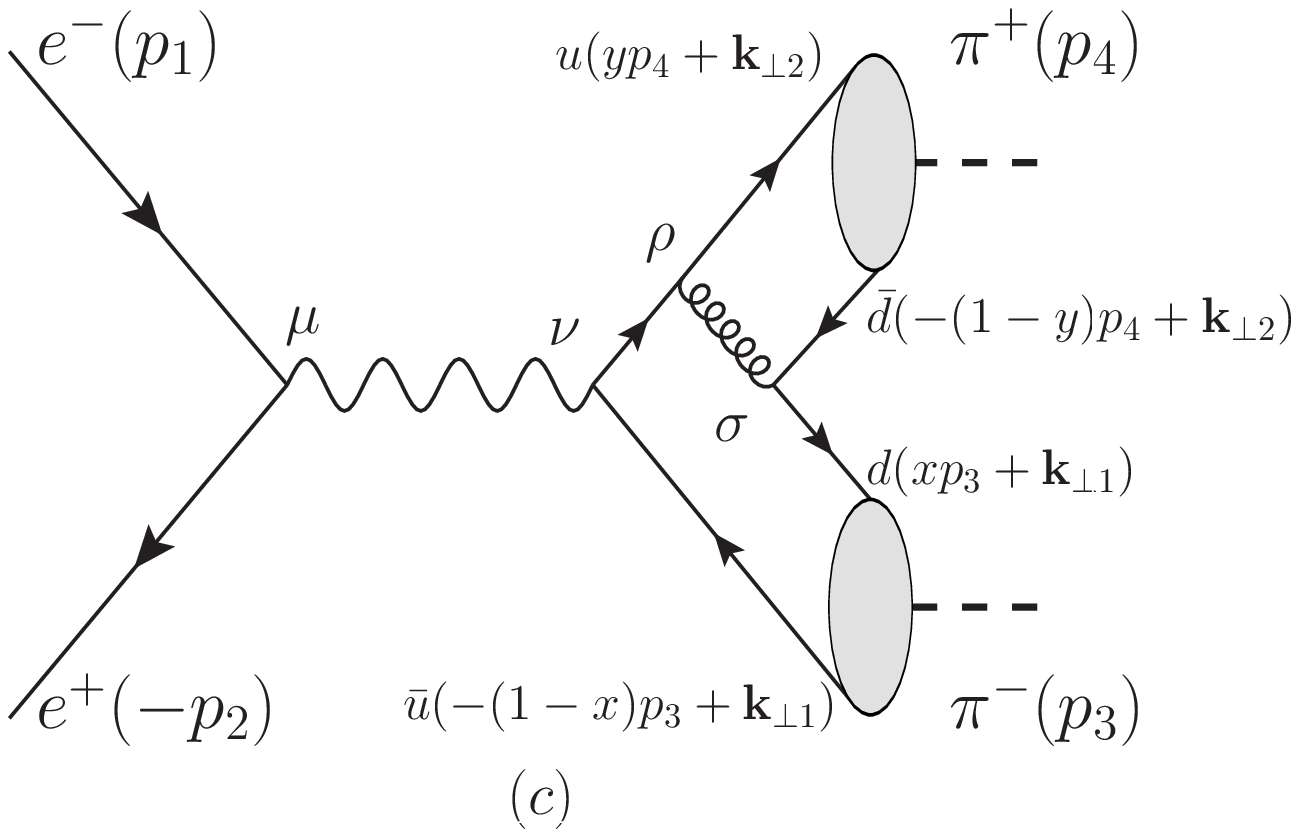}\epsfxsize 2.8 truein\epsfbox{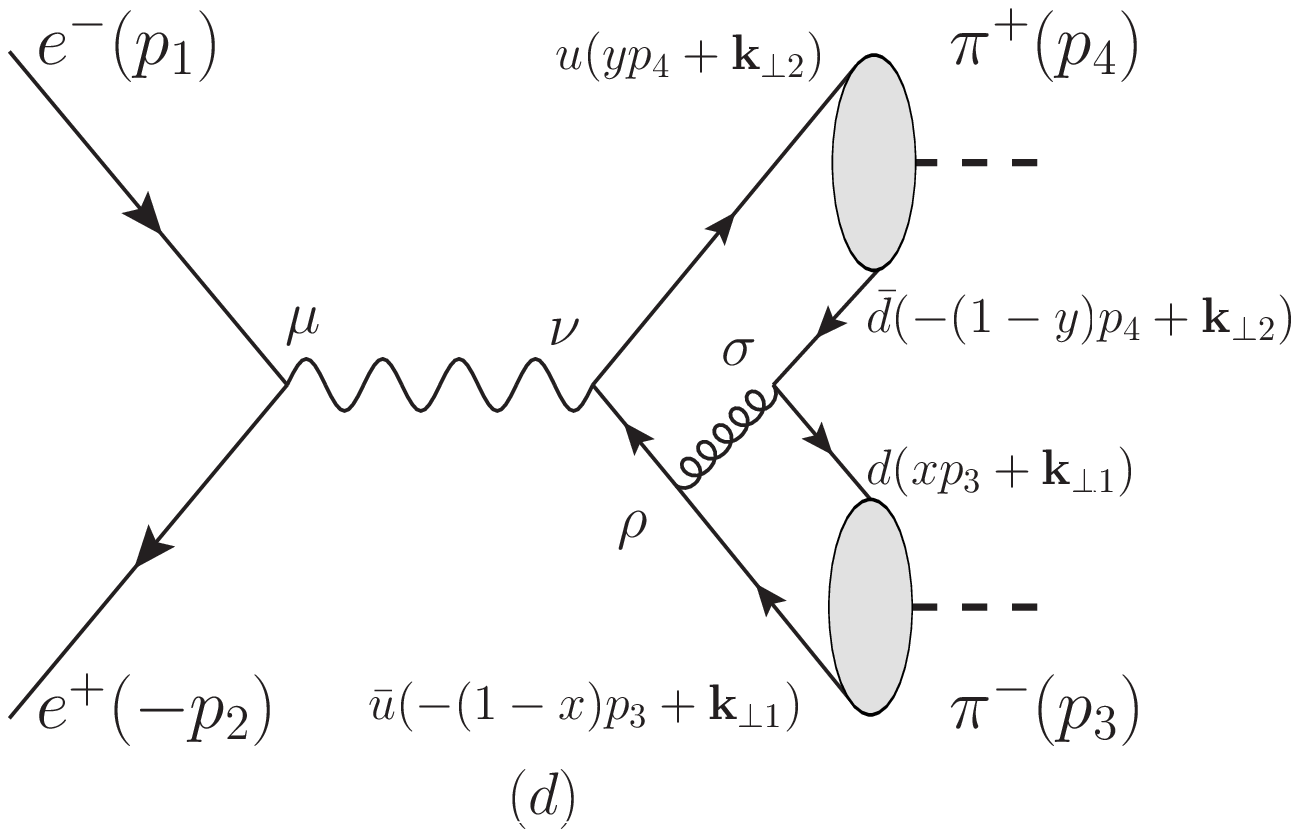}}
\caption{Diagrams for $e^{+}e^{-}\rightarrow \pi^{+}\pi^{-}$ with OPE in the leading order of pQCD.}
\label{figure:FF-OPE}
\end{figure}

In the large momentum transfer region, the perturbative QCD (pQCD) can be applied to estimate the electromagnetic form factor $F_\pi(Q^2)$ \cite{pQCD} . In the leading order of the strong interaction coupling $\alpha_s$, the corresponding Feynman diagrams are showed as Fig. \ref{figure:FF-OPE} and the corresponding contribution can be expressed as

\begin{eqnarray}
F_\pi^{(a)}(Q^2)&=&\frac{(p_4-p_3)_\nu}{-ie(p_4-p_3)^2}\int_{0}^{1}dxdy\int_{-\infty}^{\infty}d^2{\bf b}_{1}d^2{\bf b}_{2} \int_{-\infty}^{\infty}\frac{d^2{\bf k_{\perp 1}}}{(2\pi)^2}\frac{d^2{\bf k_{\perp 2}}}{(2\pi)^2}e^{-i{\bf b}_1\cdot{\bf k}_{\perp 1}-i{\bf b}_2\cdot{\bf k}_{\perp 2}} \nonumber\\
&&~~~~~~~~~~~~~~~~~~~\times e^{-S(x,y,b_1,b_2,Q)}S_t(x)S_t(y)T_H^{\nu,(a)},
\end{eqnarray}
where $b_1=|{\bf b}_1|, b_2=|{\bf b}_2|$, $S(x,y,b_1,b_2,Q)$ is the Sudakov factor in $b$ space and $S_t$ is the threshold resummation factor whose expressions can be found in \cite{Sudkov-factor-Sterman,jet-function-LiHN2002} and we also list them in the Appendix.

\begin{eqnarray}
T_H^{\nu,(a)}&=&c_f^{1\gamma} \textrm{Tr}[\Phi^{(fin)}_{\pi^+}(p_4,y,{\bf k}_{\perp 2})(-ig_s\gamma^\sigma)\Phi^{(fin)}_{\pi^-}(p_3,x,{\bf k}_{\perp 1})(-\frac{1}{3}ie\gamma^\nu)S_q(q_q) (-ig_s\gamma^\rho)] D_{\rho\sigma}(q_g),\nonumber \\
\end{eqnarray}
where $c_f^{1\gamma}=\frac{\delta_{ij}}{3}\frac{\delta_{mn}}{3}T^a_{jm}T^b_{ni}\delta_{ab}=\frac{4}{9}$ is the global color factor of the amplitude, $g_s$ is the strong coupling, $-1/3$ is the charge of $d$-quark, $e=-|e|$ is the electromagnetic coupling, $S(q_q)$ and $D_{\rho\sigma}(q_g)$ are the propagators of quark and gluon without the color indexes, $q_q$ and $q_g$  are the momenta of the  corresponding quark and gluon in the propagators with
\begin{eqnarray}
{q}_q&\equiv&[xp_3+{\bf k}_{\perp 1}]-[p_3+p_4],\nonumber\\
{q}_g&\equiv&[yp_4+{\bf k}_{\perp 2}]-[-(1-x)p_3+{\bf k}_{\perp 1}],
\end{eqnarray}
and $\Phi^{(fin)}_{\pi^{\pm}}$ are the wave functions of $\pi^{\pm}$ expressed as
\begin{eqnarray}
\Phi^{(fin)}_{\pi^+}(p_4,y,{\bf k}_{\perp 2})&=&\frac{if_\pi}{4}\Big \{\sla{{p}}_4\gamma_5\phi_{\pi}(y)
-\mu_\pi\gamma_5 \Big [\phi^P_\pi(y)-i\sigma_{\mu\nu}\Big(\frac{p_4^\mu  p_3^\nu}{p_4\cdot{p_3}}\frac{\phi^\sigma_\pi{'}(y)}{6}
-p_4^\mu\frac{\phi^\sigma_\pi(y)}{6}\frac{\partial}{\partial{\bf k}_{\perp2\nu}}\Big)\Big]\Big\},\nonumber\\
\Phi^{(fin)}_{\pi^-}(p_3,x,{\bf k}_{\perp 1})&=&\frac{if_\pi}{4}\Big\{\sla{{p}}_3\gamma_5\phi_{\pi}(x)
-\mu_\pi\gamma_5\Big[\phi^P_\pi(x)-i\sigma_{\mu\nu}\Big(\frac{p_3^\mu p_4^\nu}{p_3\cdot  p_4}\frac{\phi^\sigma_\pi{'}(x)}{6}
-p_3^\mu\frac{\phi^\sigma_\pi(x)}{6}\frac{\partial}{\partial{\bf k}_{\perp1\nu}}\Big)\Big]\Big\},\nonumber \\
\end{eqnarray}
with $f_{\pi}=0.131$GeV,

After including the contributions from the other  diagrams and some algebraic calculation, the finial expression for $F_{\pi}(Q^2)$ can be expressed as
\begin{eqnarray}
F_\pi(Q^2)&=&\int_{0}^{1}dxdy\int_{0}^{\infty}b_1db_1b_2db_2\alpha_s(\mu^2)e^{-S(x,y,b_1,b_2,Q)}S_t(x)\nonumber\\
&&\frac{16\pi f_\pi^2}{9}Q^2 \Big\{t_0+\frac{\mu_\pi^2}{Q^2}[t_1+t_2+t_3]\Big\} H_0^{(1)}(\sqrt{xy}Qb_2)\nonumber\\
&&\Big [\theta(b_1-b_2)H_0^{(1)}(\sqrt{x}Qb_1)J_0(\sqrt{x}Qb_2)+\theta(b_2-b_1)H_0^{(1)}(\sqrt{x}Qb_2)J_0^{(1)}(\sqrt{x}Qb_1)\Big ],
\label{pion-FF}
\end{eqnarray}
where the scale $\mu$ in the coupling is taken as $\textrm{max}\{\sqrt{x}Q,1/b_1,1/b_2\}$ and
\begin{eqnarray}
t_0&=&-\frac{1}{2}x\phi_\pi(y)\phi_\pi(x),\nonumber\\
t_1&=&(1-x)\phi^P_\pi(y)\phi^P_\pi(x),\nonumber\\
t_2&=&-\frac{(1+x)}{6}\phi^P_\pi(y)\phi^T_\pi(x),\nonumber\\
t_3&=&\frac{1}{3}\phi^P_\pi(y)\phi^\sigma_\pi(x).
\label{t}
\end{eqnarray}

Comparing Eq.(\ref{pion-FF},\ref{t})  with the expressions used in Ref. \cite{ZhengTaoWei2003,XingGangWu2004,Raha2010-PLB,HaoChungHu2013-PLB,ShangChen2015-PLB}, two properties of Eq.(\ref{pion-FF}) should be clarified. The first one is that Eq.(\ref{pion-FF}) is consistent with the one got by Ref. \cite{ZhengTaoWei2003} in the space-like region, the factor $1/3$ in the term $t_3$ is different with the factor $1/2$ given in Ref. \cite{XingGangWu2004}. After some careful check, we conclude this difference is due to the different deal on the term $\partial {\bf \sla{k}}_{{\bf\perp} i}/\partial{\bf k}_{\perp i\mu}$. When one takes it as $\gamma_\perp^{\mu}$ one gets $1/3$, when one takes it as $\gamma^{\mu}$ one gets $1/2$. We take the factor $1/3$ in the finial expression. In the practical numerical calculation, the contribution from this difference is very small in the space-like region and usually are neglected in some calculations, while it is not small in the time-like region and should be included. The second property of of Eq.(\ref{pion-FF}) is that there is a sign difference in the term $t_2$  between Eq.(\ref{pion-FF}) and those used in Ref. \cite{Raha2010-PLB,HaoChungHu2013-PLB,ShangChen2015-PLB}. After some check, we take Eq.(\ref{pion-FF}) as the finial result. Eq.(\ref{pion-FF}) can also be obtained via analytical continuation
of the space-like form factor \cite{ZhengTaoWei2003,XingGangWu2004} to the time-like region as the twist-2 case \cite{HaoChungHu2013-PLB}.


\section{$e^+e^-\rightarrow \pi^+\pi^-$ via two-photon-exchange}

When the TPE contributions in the process $e^+e^-\rightarrow \pi^+\pi^-$ are considered, one has the corresponding diagrams showed in Fig.3 at the leading order.

\begin{figure}[htbp]
\center{\epsfxsize 2.8 truein\epsfbox{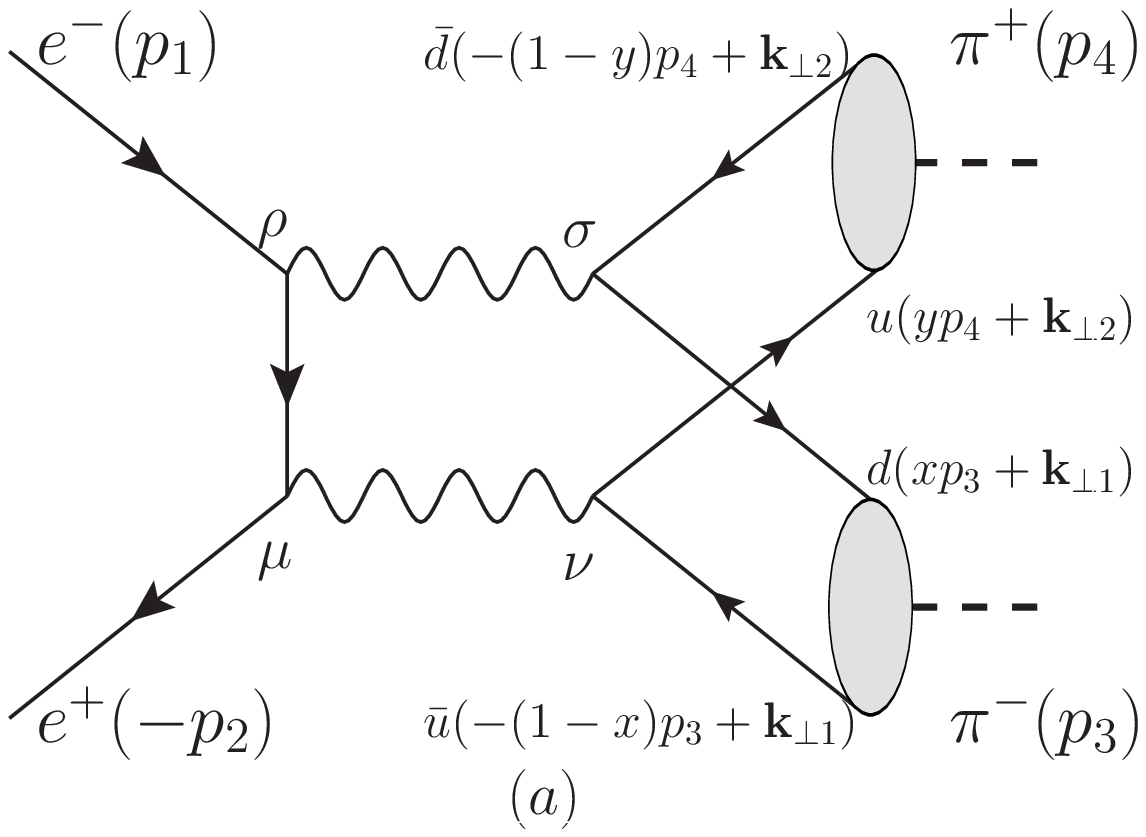}\epsfxsize 2.8 truein\epsfbox{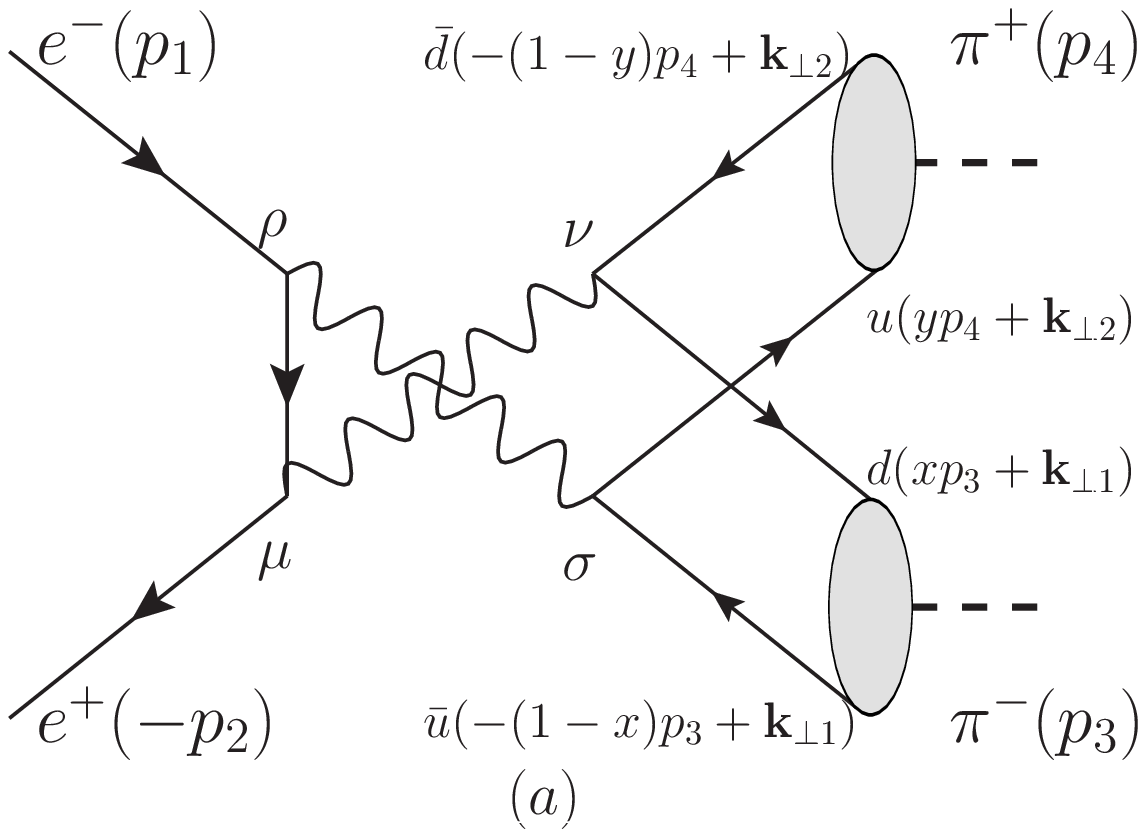}}
\caption{Diagrams for TPE for $e^{+}e^{-}\rightarrow \pi^{+}\pi^{-}$  with two-photon exchange (TPE) in the leading order of pQCD.}
\label{figure:Amp-TPE}
\end{figure}
The  amplitude  corresponding to Fig. \ref{figure:Amp-TPE}(a) can be expressed as
\begin{eqnarray}
i\mathcal{M}^{2\gamma,(a)}&=&\int dxdy \int d^2{\bf b}_1d^2{\bf b}_2\int \frac{d^2{\bf k}_{\perp 1}}{(2\pi)^2}\frac{d^2{\bf k}_{\perp 2}}{(2\pi)^2}e^{-i{\bf b}_1\cdot{\bf k}_{\perp 1}-i{\bf b}_2\cdot{\bf k}_{\perp 2}} \nonumber\\
&&~~~\times e^{-S(x,y,b_1,b_2,Q)}T_H^{2\gamma,(a)} \nonumber \\
&\triangleq &\int K\ast T_H^{2\gamma,(a)},
\end{eqnarray}
where
\begin{eqnarray}
T_H^{2\gamma,(a)}&=&\bar{u}(-{p}_2,s_2)(-ie\gamma^\mu)S_e(q_e)(-ie\gamma^\rho)u({p}_1,s_1)D_{\rho\sigma}(q_1)D_{\mu\nu}(q_2)\nonumber\\
&& c_{2\gamma}\textrm{Tr}[\Phi _{\pi}^{(f)}(p_4,y,{\bf b}_2)(\frac{2}{3}ie\gamma^\nu)\Phi_{mn,\pi}^{(f)}(p_3,x,{\bf b}_1)(-\frac{1}{3}ie\gamma^\sigma)]\nonumber\\
& \triangleq & \bar{u}(-{p}_2,s_2)\gamma^\mu\gamma^\omega\gamma^\rho u({p}_1,s_1)q_{e,\omega}T^{(a)}_{\mu\rho}(Q^2,\theta,{\bf b}_1,{\bf k}_{\perp 1},{\bf b}_2,{\bf k}_{\perp 2}),
\end{eqnarray}
with $c_{2\gamma}=\frac{\delta_{ij}}{3}\frac{\delta_{ij}}{3}=\frac{1}{3}$ the global color factor and the momenta in the propagators
\begin{eqnarray}
{q}_e&=&-{p}_2+q_2,\nonumber\\
{q}_1&=&[xp_3+{\bf k}_{\perp 1}]-[-(1-y)p_4+{\bf k}_{\perp 2}],\nonumber\\
{q}_2&=&[yp_4+{\bf k}_{\perp 2}]-[-(1-x)p_3+{\bf k}_{\perp 1}],
\end{eqnarray}
and
\begin{eqnarray}
&&T^{(a)}_{\mu\rho}(Q^2,\theta,{\bf b}_1,{\bf k}_{\perp 1},{\bf b}_2,{\bf k}_{\perp 2}) \nonumber \\
& = &c_{2\gamma} \textrm{Tr}[\Phi_\pi(p_4,y,{\bf b}_2)(\frac{2}{3}ie\gamma_\mu)\Phi_\pi(p_3,x,{\bf b}_1)(-\frac{1}{3}ie\gamma_\rho)](-ie)^2\frac{-i}{q_1^2+i\epsilon}\frac{-i}{q_2^2+i\epsilon}\frac{i}{q_e^2+i\epsilon}.
\label{T_munu}
\end{eqnarray}
Using the relation
\begin{eqnarray}
\gamma^\mu\gamma^\omega\gamma^\rho=
g^{\mu\omega}\gamma^\rho-g^{\mu\rho}\gamma^\omega+g^{\omega\rho}\gamma^\mu-i\gamma^5\epsilon^{\mu\omega\rho\sigma}\gamma_\sigma,
\end{eqnarray}
the amplitude $i\mathcal{M}^{2\gamma,(a)}$ can be expressed in a similar form as $i\mathcal{M}^{1\gamma}$ and one has

\begin{eqnarray}
&&T_H^{2\gamma,(a)}({p}_1,s_1;{p}_2,s_2;p_3,p_4) \nonumber \\
&=&\bar{u}(-{p}_2,s_2)\gamma^\rho u({p}_1,s_1)q_{e,\omega}T^{(a)}_{\omega\rho}-\bar{u}(-{p}_2,s_2)\gamma^\omega u({p}_1,s_1)q_{e,\omega}T^{(a)}_{\mu\mu}\nonumber \\
&&+\bar{u}(-{p}_2,s_2)\gamma^\mu u({p}_1,s_1)q_{e,\rho}T^{(a)}_{\mu\rho} -\bar{u}(-{p}_2,s_2)\gamma^\sigma u({p}_1,s_1)i\gamma_5\epsilon^{\mu\omega\rho}_{~~~~\sigma}q_{e,\omega}T^{(a)}_{\mu\rho}) \nonumber \\
&=& [{\bar u}(-p_2,m_e)\gamma^\mu u(p_1,m_e)][q_{e,\omega}T^{(a)}_{\omega\mu}-q_{e,\mu}T^{(a)}_{\rho\rho}+q_{e,\rho}T^{(a)}_{\mu\rho}]\nonumber \\
&&+[{\bar u}(-p_2,m_e)\gamma_5\gamma^\mu u(p_1,m_e)][-i\epsilon^{\sigma\omega\rho}_{~~~~\mu}q_{e,\omega}T^{(a)}_{\sigma\rho}],\nonumber \\
&\triangleq& [{\bar u}(-p_2,m_e)(-ie\gamma_\mu) u(p_1,m_e)]D^{\mu\nu}(q) T_{\nu}^{(a),eff}\nonumber \\
&&+[{\bar u}(-p_2,m_e)(-ie\gamma_5\gamma_\mu) u(p_1,m_e)]D^{\mu\nu}(q) \bar{T}_{\nu}^{(a),eff},
\end{eqnarray}
with
\begin{eqnarray}
T_{\nu}^{(a),eff}&=&\frac{1}{-ie}\frac{q^2}{-i}[q_{e,\omega}T^{(a)}_{\omega\nu}-q_{e,\nu}T^{(a)}_{\rho\rho}+q_{e,\rho}T^{(a)}_{\nu\rho}],\nonumber \\
\bar{T}_{\nu}^{(a),eff}&=&\frac{1}{-ie}\frac{q^2}{-i}[-i\epsilon^{\sigma\omega\rho}_{~~~~\nu}q_{e,\omega}T^{(a)}_{\sigma\rho}].
\label{T_nu_eff}
\end{eqnarray}
Generally, $T_{\nu}^{(a),eff}$ can be written as $c_1p_{1\nu}+c_2p_{2\nu}+c_3p_{3\nu}$, using the approximation $m_e=0$, the first two terms give no contributions and one get  $T_{\nu}^{(a),eff} \propto (p_4-p_3)_\nu$ and finally
\begin{eqnarray}
&&i\mathcal{M}^{2\gamma,(a)} \nonumber \\
&=&\int K \ast [{\bar u}(-p_2,m_e)(-ie\gamma_\mu) u(p_1,m_e)]D^{\mu\nu}(q)T_{\nu}^{eff}] \nonumber \\
&&+\int K \ast [{\bar u}(-p_2,m_e)(-ie\gamma_5\gamma_\mu) u(p_1,m_e)]D^{\mu\nu}(q)\tilde{T}_{\nu}^{eff,(a)}]\nonumber \\
&\triangleq& [{\bar u}(-p_2,m_e)(-ie\gamma_\mu) u(p_1,m_e)]D^{\mu\nu}(q)[-ie (p_4-p_3)_\nu \tilde{F}_\pi^{(a)}(Q^2,\theta)],\nonumber \\
&&+[{\bar u}(-p_2,m_e)(-ie\gamma_5\gamma_\mu) u(p_1,m_e)]D^{\mu\nu}(q)[-ie (p_4-p_3)_\nu \tilde{G}_\pi^{(a)}(Q^2,\theta)],
\end{eqnarray}
where $\tilde{F}_\pi(Q^2,\theta),\tilde{G}_\pi(Q^2,\theta)$ are expressed as
\begin{eqnarray}
\label{FFs of TPE part1}
\tilde{F}_\pi^{(a)}(Q^2,\theta)&=& \int \frac{(p_4-p_3)^\nu}{-ie(p_4-p_3)^2}T_\nu^{(a),eff},\nonumber \\
\tilde{G}_\pi^{(a)}(Q^2,\theta)&=& \int \frac{(p_4-p_3)^\nu}{-ie(p_4-p_3)^2}\bar{T}_\nu^{(a),eff}.
\label{FFs of TPE}
\end{eqnarray}
The contribution from the Fig. \ref{figure:Amp-TPE} (b) can be also get in a similar way. Due to the similar form with $F_{\pi}(Q^2)$, we call $\tilde{F}_\pi^{(a)}(Q^2,\theta),\tilde{G}_\pi^{(a)}(Q^2,\theta)$ as the general form factors in the following and the finial expressions for the general form factors can be get from the Eq. (\ref{T_munu},\ref{T_nu_eff},\ref{FFs of TPE}).

After some calculation, one has
\begin{eqnarray}
\tilde{F}_\pi(Q^2,\theta)& \triangleq &\tilde{F}^{(a)}_\pi(Q^2,\theta)+\tilde{F}^{(b)}_\pi(Q^2,\theta),\nonumber\\
\tilde{F}^{(b)}_\pi(Q^2,\theta)&=&-\tilde{F}^{(a)}_\pi(Q^2,\theta+\pi),
\end{eqnarray}
where
\begin{eqnarray}
\tilde{F}^{(a)}_\pi(Q^2,\theta)&=&\frac{c_{2\gamma} e^2 f_\pi^2 Q^2}{36\pi}\int b_2db_2\int dxdy~e^{-S(x,y,b_1,b_2,Q)}\nonumber\\
&&\times\bigg\{\frac{1}{2}\phi_\pi(x)\phi_\pi(y)Q^2(\!-\!\cos\theta\!+\!x\!+\!y\!-\!1) +\mu_\pi^2\big[\phi^P_\pi(x)\phi^P_\pi(y)(\!-\!\cos\theta\!+\!x\!+\!y\!-\!1)\nonumber\\
&&-\frac{1}{36}\phi^T_\pi(x)\phi^T_\pi(y)(\!-\!\cos\theta\!+\!x\!+\!y\!-\!1)+\frac{1}{24}\phi^T_\pi(x)\phi^\sigma_\pi(y) +\frac{1}{24}\phi^\sigma_\pi(x)\phi^T_\pi(y)\big]\bigg\}\nonumber\\
&&\times \tilde{H}(x,y,Q,b_2,\theta),
\end{eqnarray}
and
\begin{eqnarray}
\tilde{H}(x,y,Q,b_2,\theta)&=&\int d\phi_{b_2} dk_{\perp 3x}e^{-ib_{2x}k_{\perp 3x}}\nonumber\\
&&\times\bigg\{\frac{2\sqrt{2}e^{\frac{|b_{2y}|}{\sqrt{2}}\left(-\sqrt{P_1^{(1)}(x,y,Q,k_{\perp3x},\theta)-i\epsilon}\right)}}{\sqrt{P_1^{(1)}\!(x,y,\!Q,k_{\perp3x},\!\theta)\!-\!i\epsilon} P_2^{(1)}\!(x,y,\!Q,k_{\perp3x},\!\theta)P_3^{(1)}\!(x,y,\!Q,k_{\perp3x},\!\theta)}\nonumber\\
&&-\frac{e^{|b_{2y}|\left(-\sqrt{P_1^{(2)}(x,y,Q,k_{\perp3x})-i\epsilon}\right)}}{\sqrt{P_1^{(2)}(x,y,Q,k_{\perp3x})-i\epsilon} P_2^{(2)}(x,y,Q)P_3^{(2)}(x,y,Q,k_{\perp3x},\theta)}\nonumber\\ &&+\frac{e^{|b_{2y}|\left(-\sqrt{P_1^{(3)}(x,y,Q,k_{\perp3x})-i\epsilon}\right)}}{\sqrt{P_1^{(3)}(x,y,Q,k_{\perp3x})-i\epsilon} P_2^{(3)}(x,y,Q)P_3^{(3)}(x,y,Q,k_{\perp3x},\theta)}\bigg\},
\end{eqnarray}
with $b_{2y} \triangleq b_2 \sin{\phi_{b_2}}$, $b_{2x}\triangleq b_2 \cos{\phi_{b_2}}$, $k_{\perp 3}=k_{\perp 2}-k_{\perp 1}=\{k_{\perp 3x},k_{\perp 3y}\}$, $\epsilon=0^+$ and
\begin{eqnarray}
&&P_1^{(1)}(x,y,Q,k_{\perp3x},\theta)=2 k_{\perp3x}^2+2 k_{\perp3x}Q\sin\theta+Q^2(-\cos\theta(x+y-1)+2xy-x-y+1)\nonumber\\
&&+2m_e^2,\nonumber\\
&&P_2^{(1)}(x,y,Q,k_{\perp3x},\theta)=2 k_{\perp3x}Q\sin\theta+Q^2(-\cos\theta(x+y-1)+x-y+1)+2m_e^2,\nonumber\\
&&P_3^{(1)}(x,y,Q,k_{\perp3x},\theta)=2 k_{\perp3x} Q\sin\theta+Q^2(-\cos\theta(x+y-1)-x+y+1)+2m_e^2,\nonumber\\
&&P_1^{(2)}(x,y,Q,k_{\perp3x})=k_{\perp3x}^2+Q^2(x-1)y,\nonumber\\
&&P_2^{(2)}(x,y,Q)=Q^2 (x-y),\nonumber\\
&&P_3^{(2)}(x,y,Q,k_{\perp3x},\theta)=P_3^{(1)}(x,y,Q,k_{\perp3x},\theta),\nonumber\\
&&P_1^{(3)}(x,y,Q,k_{\perp3x})=k_{\perp3x}^2+Q^2x(y-1),\nonumber\\
&&P_2^{(3)}(x,y,Q)=P_2^{(2)}(x,y,Q),\nonumber\\
&&P_3^{(3)}(x,y,Q,k_{\perp3x},\theta)=P_2^{(1)}(x,y,Q,k_{\perp3x},\theta).
\end{eqnarray}

Furthermore, the cross section from the interference of $\mathcal{M}^{2\gamma}$ and $\mathcal{M}^{1\gamma}$ can expressed as
\begin{eqnarray}
d\sigma_{un}^{2\gamma}
&=&\frac{1}{2}e^2\ sin^2\theta\{2Re[F^*_\pi(Q^2)\tilde{F}_\pi(Q^2,\theta)]\},
\end{eqnarray}
and there is no contribution from $\tilde{G}_\pi(Q^2,\theta)$.

\section{The input }
In the time-like region, in principle the contributions from the resonances should also be considered. In this work, we limit our discussion at the high energy region and focus on the TPE effects, so we neglect the contributions from the resonances at present and the needed input are the same as those used in space-like region. For simplicity, we directly take $n_f=3,\Lambda=0.2$GeV in the Sudakov factor and neglect the dependence of $n_f$ and $\Lambda$ on $Q^2, 1/b_1$ and $1/b_2$. All other inputs are taken as same as those used in Ref. \cite{HaoChungHu2013-PLB} which means the asymptotic  two-parton twist-2 and twist-3 DAs are taken
\begin{eqnarray}
\phi_{\pi}(x)&=&6x(1-x)[1+a_2C^{3/2}_{2}(1-2x)], \nonumber \\
\phi^{P}_{\pi}(x)&=&1, \nonumber \\
\phi^{\sigma}_{\pi}(x)&=&6x(1-x),  \nonumber \\
\phi^{T}_{\pi}(x)&=&d\phi^{\sigma}_{\pi}(x)/dx =6(1-2x),
\end{eqnarray}
with $a_2=0.2$ and the Gegenbauer polynomial $C^{3/2}_2(u)=(3/2)(5u^2-1)$. The normalization of the above DAs is a little different with that in Ref. \cite{HaoChungHu2013-PLB}. The associated chiral scale is taken as $\mu_\pi=1.3$GeV, the shape parameter in the threshold resummation factor $S_t(x)$  is taken as $c=0.4$ and the renormalization scale used in the $\alpha_{S}$ and Sudakov factor is taken as $\mu=max(\sqrt{x}Q,1/b_1,1/b_2)$.

Other forms of DAs are also used for estimation and the practical numerical results show the form factors are a little sensitive on the input DAs.  Since our focus is on the TPE effects in $e^+e^-\rightarrow \pi^+\pi^-$, we do not go to discuss the detail of the  dependence of the pion form factor $Q^2|F_{\pi}(Q^2)|$ on the input DAs.


\section{Numerical results and discussion}
Using the inputs suggested in the last section,  the form factors $F_{\pi}(Q^2),\tilde{F}_\pi(Q^2,\theta)$ can be calculated directly by the  numerical method. In our numerical calculation, we use the function NIntegrate in the Mathematica to do the integration and also the Bessel function in the Mathematica are used directly. The function Vegas in the package Cuba \cite{Cuba} is also used to check the numerical calculation and we find it gives the same result. We want to point out that the integration include the Bessel function should be dealt carefully. The integration of $Q^2|\tilde{F}_\pi(Q^2,\theta)|$ is heavy and in the practical calculation, we at first calculate the results at some points with the relative precision about $1\%$ and then fit the results.

\begin{figure}[!htbp]
\center{\epsfxsize 3.4 truein\epsfbox{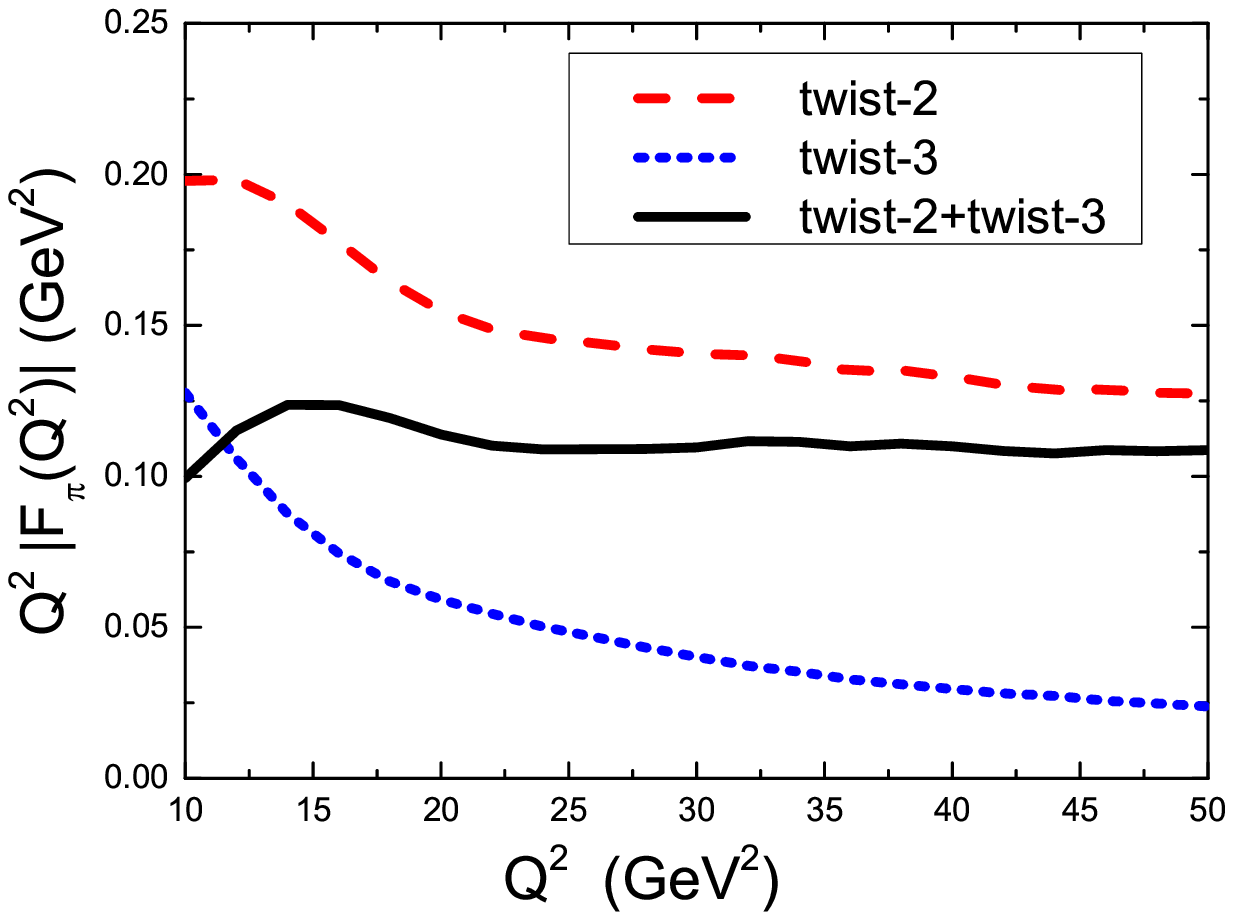}\epsfxsize 3.4 truein\epsfbox{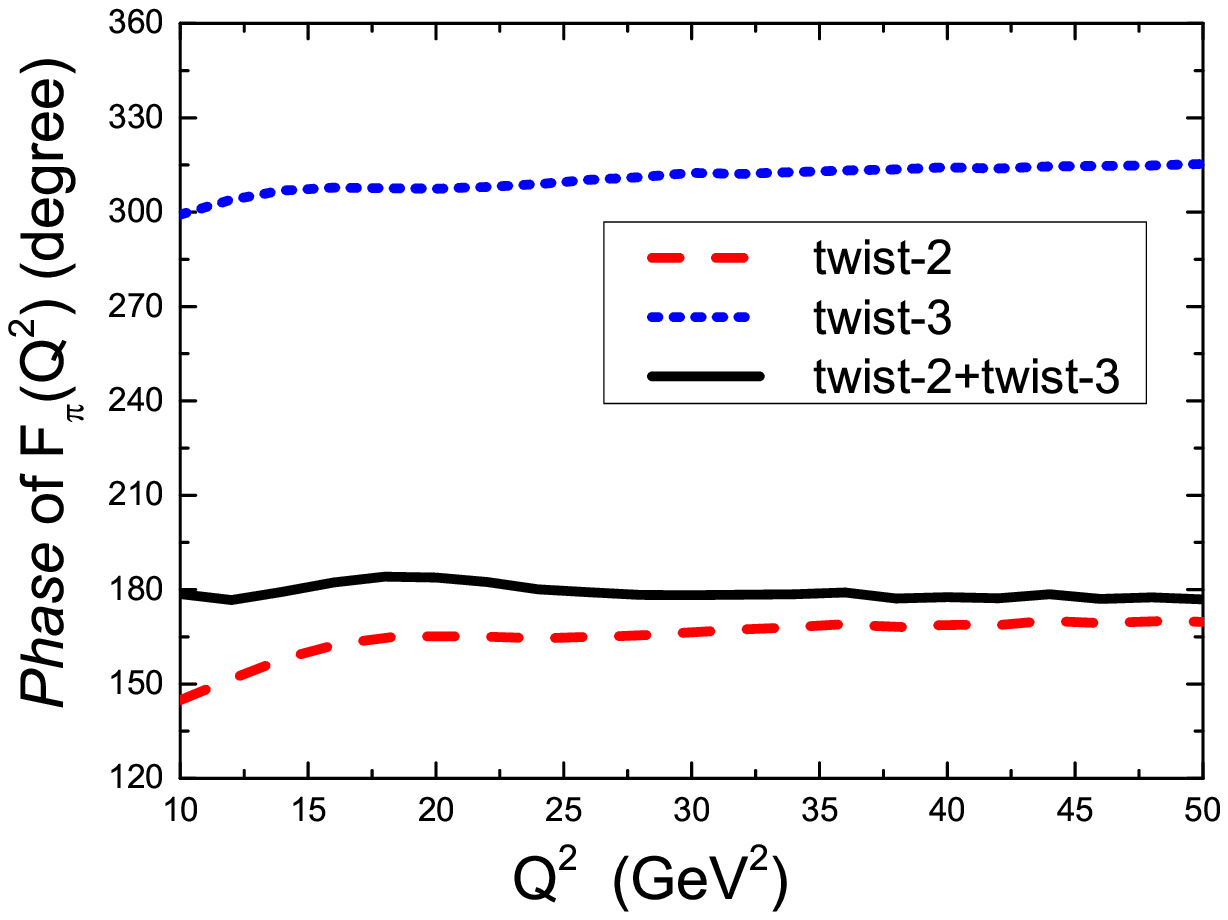}}
\caption{Results for $Q^2 F_\pi(Q^2)$ {\it vs} . $Q^2$. The left panel is the result for $Q^2 |F_\pi(Q^2)|$ {\it vs} . $Q^2$ and the right panel is the result for  the phase of $F_\pi(Q^2)$ {\it vs} . $Q^2$. }
\label{figure:OPE-full}
\end{figure}

\begin{figure}[!htbp]
\center{\epsfxsize 3.4 truein\epsfbox{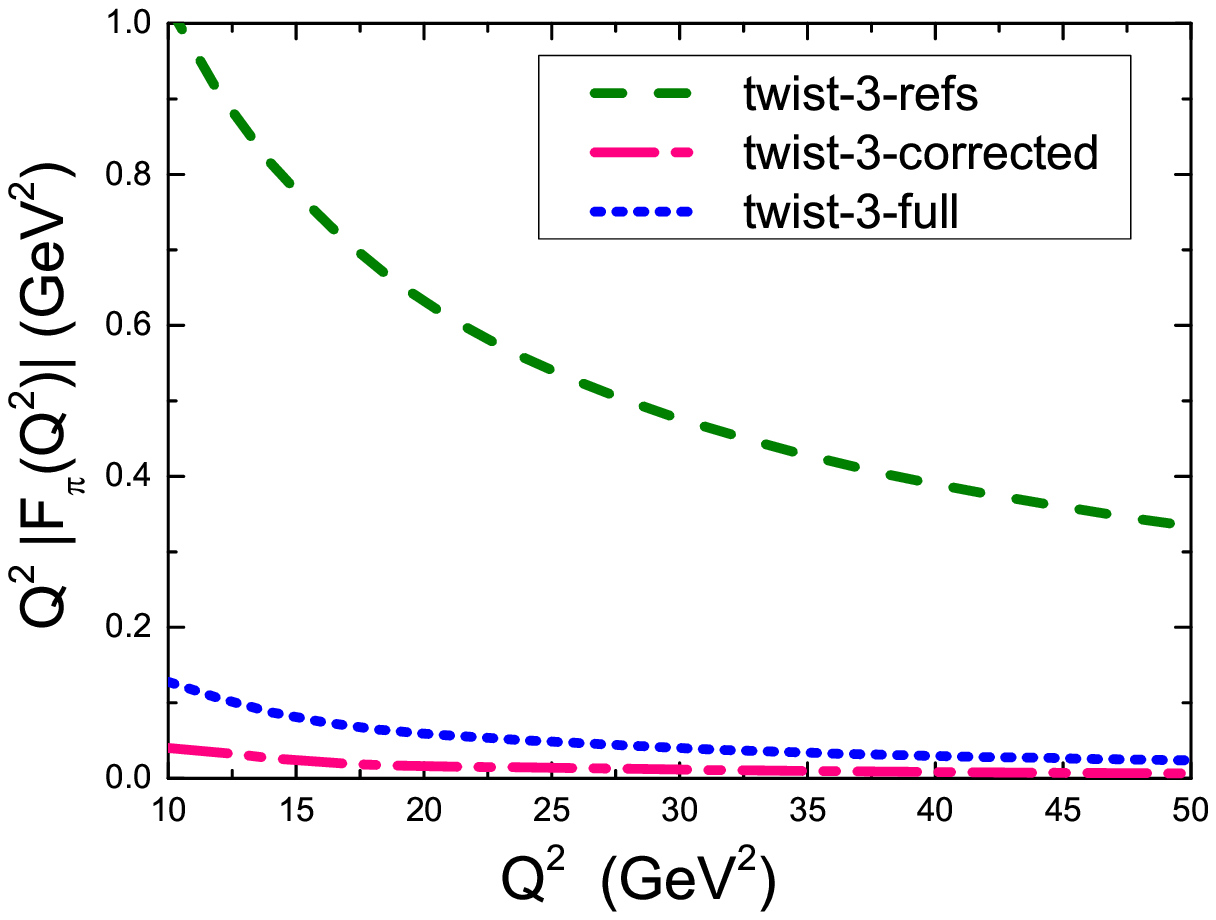}\epsfxsize 3.4 truein\epsfbox{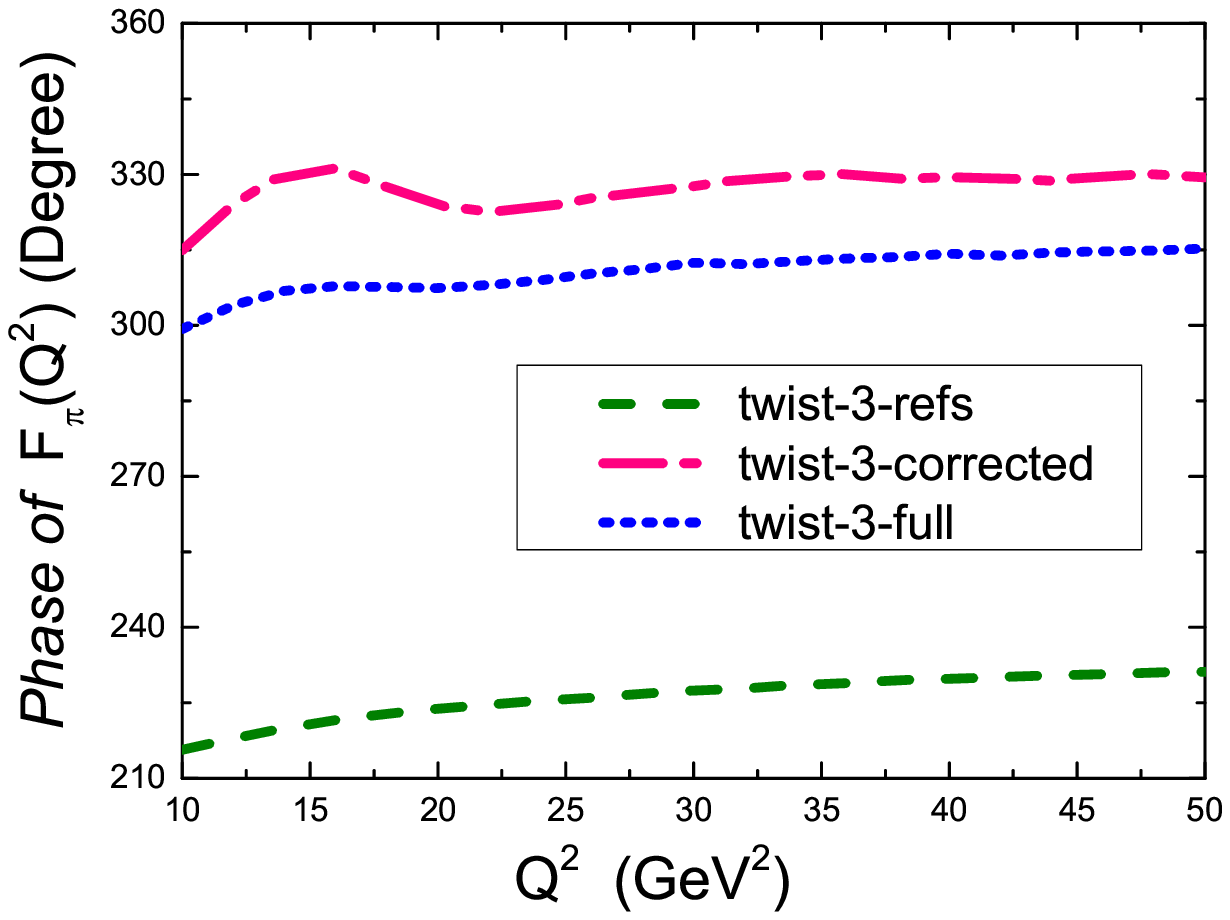}}
\caption{Comparison of the contributions from twist-3 DAs to $Q^2 |F_\pi(Q^2)|$ and the phase of  $F_\pi(Q^2)$ with different expressions. The  olive dashed curves labelled as ``twist-3-refs" refers to the results by replacing $t_1+t_2+t_3$ in Eq. (\ref{pion-FF}) with $t_1-t_2$ which was given in \cite{Raha2010-PLB} and then used in Ref.  \cite{HaoChungHu2013-PLB,ShangChen2015-PLB}, the pink dashed-doted curves labelled as ``twist-3-corrected" refers to the results by replacing $t_1+t_2+t_3$ in Eq. (\ref{pion-FF}) with $t_1+t_2$ and the  black solid curves labelled as ``twist-3-full" refers to the results by Eq. (\ref{pion-FF}).}
\label{figure:OPE-twist3-comparison}
\end{figure}

The numerical results for $Q^2|F_{\pi}(Q^2)|$ and the phase of $F_{\pi}(Q^2)$ are presented in Fig. \ref{figure:OPE-full}. The red dashed curves refer to the contribution from twist-2 DA, the blue dotted curves refer to the contribution from twist-3 DAs and the black solid curves refer to the contribution from their sum. The contribution from the twist-2 DA is almost same with that presented in \cite{HaoChungHu2013-PLB}. The contribution from the twist-3 DAs is much smaller than that from the twist-2 DA, which is very different with the property presented in \cite{HaoChungHu2013-PLB,ShangChen2015-PLB}. For comparison, three results are presented in Fig. \ref{figure:OPE-twist3-comparison} to show the reason of the large difference.

In Fig. \ref{figure:OPE-twist3-comparison}, the olive dashed curves labelled as ``twist-3-refs" refer to the results by replacing $t_1+t_2+t_3$ in Eq. (\ref{pion-FF}) with $t_1-t_2$  which was given in \cite{Raha2010-PLB} and then used in Ref.  \cite{HaoChungHu2013-PLB,ShangChen2015-PLB}, the pink dashed-doted curves labelled as ``twist-3-corrected" refers to the results by replacing $t_1+t_2+t_3$ in Eq. (\ref{pion-FF}) with $t_1+t_2$ and the black solid curves labelled as ``twist-3-full" refer to the results from Eq. (\ref{pion-FF}). The numerical results ``twist-3-Refs" are almost same with the corresponding results in Fig.5 of Ref. \cite{HaoChungHu2013-PLB}. The comparison of the results ``twist-3-refs" and ``twist-3-corrected" shows that there is large cancellation between the contributions from the terms $t_1$ and $t_2$. The comparison of the results ``twist-3-corrected" and ``twist-3-full" shows the contribution from the term $t_3$ is also important. The property of the contribution from the term $t_3$ is very different with that in the space-like region where the contribution from this term is small.

\begin{figure}[!htbp]
\center{\epsfxsize 6 truein\epsfbox{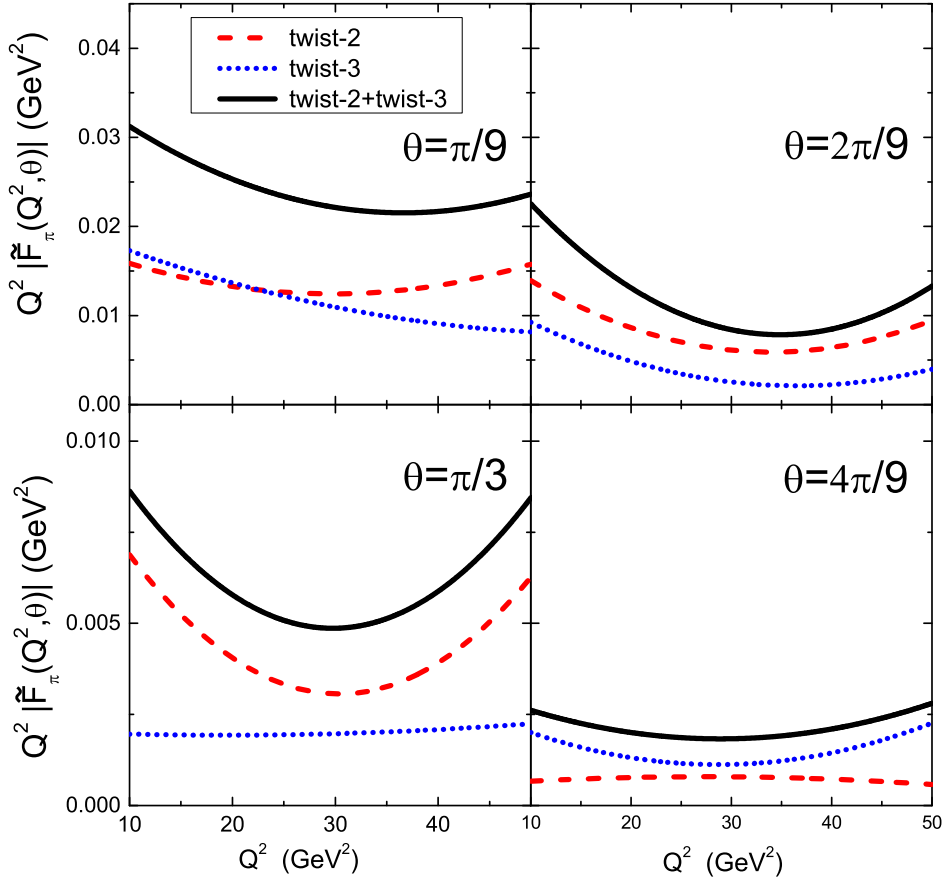}}
\caption{The numerical results for $Q^2|\tilde{F}_\pi(Q^2,\theta)|$ {\it vs.} $Q^2$ at $\theta=(1/9,2/9,1/3,4/9)\pi$ from twist-2 DA (red dashed), twist-3 DAs (blue dotted) and their sum (black solid), respectively.}
\label{figure:FFbar-abs-vs-QQ}
\end{figure}

\begin{figure}[htbp]
\center{\epsfxsize 6 truein\epsfbox{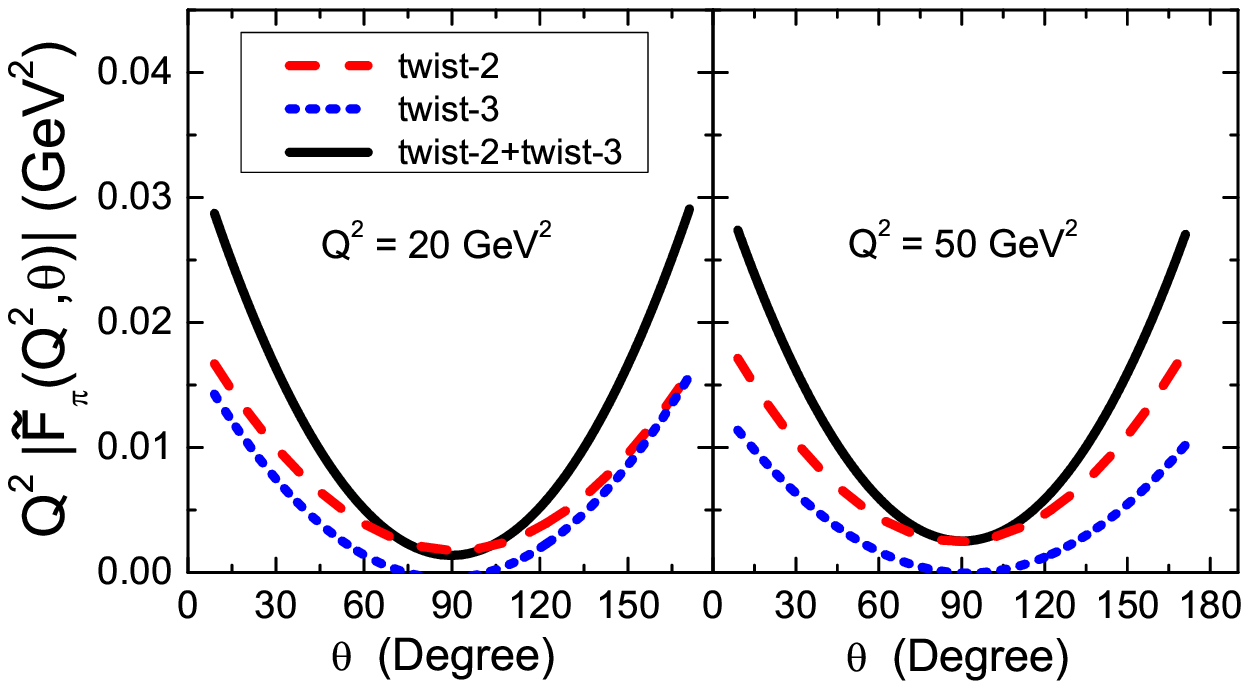}}
\caption{The numerical results for $Q^2|\tilde{F}_\pi(Q^2,\theta)|$ {\it vs.} $\theta$ at $Q^2=(20,50)$ GeV$^2$  from twist-2 DA (red dashed), twist-3 DA (blue dotted) and their sum (black solid), respectively.}
\label{figure:FFbar-abs-vs-theta}
\end{figure}

The numerical results for $Q^2|\tilde{F}_\pi(Q^2,\theta)|$ {\it vs.} $Q^2$ at $\theta =(1/9,2/9,1/3,4/9)\pi$ are presented in Fig. \ref{figure:FFbar-abs-vs-QQ}. The red dashed curves refer to the contribution from twist-2 DA, the blue dotted curves refer to the contribution from twist-3 DAs and the black solid curves refer to the contribution from their sum. One can see the magnitudes of $Q^2|\tilde{F}_\pi(Q^2,\theta)|$ are about ($10\%-20\%$) of $Q^2|F_\pi(Q^2)|$ at small $\theta$ which means the absolute contributions from the TPE effects are not small. This is natural since naively the ratio is expected as $\alpha_{QED}/\alpha_S$ due to Fig. \ref{figure:FF-OPE} and Fig. \ref{figure:Amp-TPE}. This property is differen with the TPE corrections in the elastic $ep$ scattering at small momentum transfer where the relative corrections are expected as $\alpha_{QED}$.   The $Q^2|\tilde{F}_\pi(Q^2,\theta)|$ also shows strong angle dependence which is the most interesting property different with $F_{\pi}(Q^2)$. The manifest dependence of $Q^2|\tilde{F}_\pi(Q^2,\theta)|$ on $\theta$ at $Q^2=(20,50)$ GeV$^2$ are presented in Fig. \ref{figure:FFbar-abs-vs-theta}.

\begin{figure}[htbp]
\center{\epsfxsize 7.5 truein\epsfbox{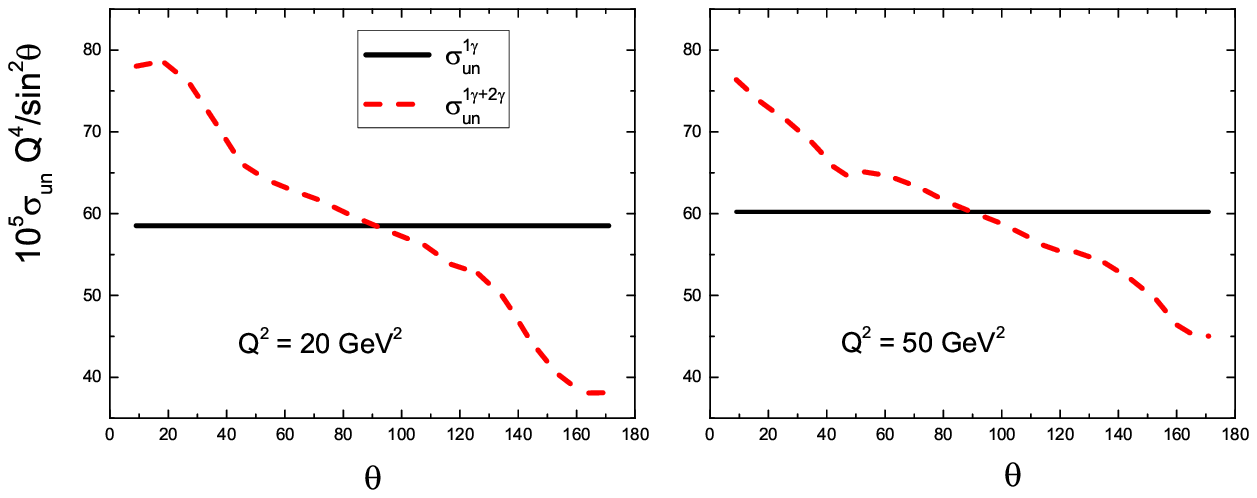}}
\caption{The numerical results for $d\sigma_{un}Q^4/sin^2\theta$ {\it vs.} $\theta$ at $Q^2=(20,50)$ GeV$^2$ from the OPE (black solid) and OPE+TPE (red dashed), respectively.}
\label{figure:cross-section}
\end{figure}

The normalized cross sections $d\sigma_{un}Q^4/sin^2\theta$ from the OPE (black solid curves) and OPE+TPE (red dashed curves) are presented in Fig. \ref{figure:cross-section}, where one can see a manifest asymmetry in the angle dependence of the cross section after including the TPE effects. The existing of such asymmetry is a direct single of the TPE effects. The measurements of such asymmetry can help us understand the TPE effects.

In summary, in this work the TPE effects in the process $e^+e^- \rightarrow \pi^+\pi^-$ at large momentum transfer are discussed  within the perturbative QCD (pQCD). The TPE contributions to the cross section are calculated and we find the asymmetry of the differential cross section on the scattering angle reaches about $10\%-20\%$ at small angle. The time-like electromagnetic form factor of pion at the leading order of $\alpha_S$ from the twist-3 DAs is also discussed and the comparison of our results with those in the references are presented.

\section{Acknowledgments}
The author Hai-Qing Zhou would like to thank Hsiang-nan Li, Xing-Gang Wu and Shan Cheng for their kind and helpful discussions. This work is supported by the  National Natural Science Foundations of China under Grant No. 11375044.
\section{Appendix}
In this Appendix, some expressions used in the practical calculation are listed.

The Sudkov factor $S(x,y,b_1,b_2,Q)$ \cite{Sudkov-factor-Sterman} is expressed as
\begin{eqnarray}
S(x,y,b_1,b_2,Q)&=&s(xQ,b_1)+s(yQ,b_2)+s((1-x)Q,b_1)+s((1-y)Q,b_2)\nonumber\\
&&-\frac{1}{\beta_0}\text{ln}\left(\frac{\hat{t}}{-\hat{b}_1}\right) -\frac{1}{\beta_0}\text{ln}\left(\frac{\hat{t}}{-\hat{b}_2}\right),
\end{eqnarray}
where
\begin{eqnarray}
s(xQ,1/b)&=&\frac{A^{(1)}}{2\beta_0}\hat{q}\text{ln} \left(\frac{\hat{q}}{-\hat{b}}\right) +\frac{A^{(2)}}{4\beta_0^2} \left(\frac{\hat{q}}{-\hat{b}}-1\right) -\frac{A^{(1)}}{2\beta_0}(\hat{b}+\hat{q})\nonumber\\
&&-\frac{4A^{(1)}\beta_1}{16\beta_0^3}\hat{q}
\left[\frac{1+\text{ln}(-2\hat{b})}{-\hat{b}} -\frac{1+\text{ln}(2\hat{q})}{\hat{q}}\right]\nonumber\\
&&-\left[\frac{A^{(2)}}{4\beta_0^2} -\frac{A^{(1)}}{4\beta_0} \text{ln}\left(\frac{1}{2}e^{2\gamma_E-1}\right)\right]\text{ln}\left(\frac{\hat{q}}{-\hat{b}}\right)\nonumber\\
&&-\frac{4A^{(1)}\beta_1}{32\beta_0^3} \left[\text{ln}^2(-2\hat{b})-\text{ln}^2(2\hat{q})\right],
\end{eqnarray}
with
\begin{eqnarray}
&&\hat{t}=ln(\frac{t}{\Lambda_{QCD}}), \qquad t=max(\sqrt{x}Q,1/b_1,1/b_2),\nonumber\\
&&\hat{b}=ln(b\Lambda_{QCD}),\qquad \hat{q}=ln[\frac{xQ}{\sqrt{2}\Lambda_{QCD}}],\nonumber\\
&&A^{(1)}=C_F=\frac{4}{3},\nonumber\\
&&A^{(2)}=(\frac{67}{27}-\frac{\pi^2}{9})N_c-\frac{10}{27}N_f+\frac{8}{3}\beta_0ln(\frac{e^{\gamma_E}}{2}),\nonumber\\
&&\beta_0=\frac{11N_c-2N_f}{12}=\frac{9}{4},\qquad \beta_1=\frac{51N_c-19N_f}{24}=4,\nonumber\\
&& N_c=N_f=3.
\end{eqnarray}

The jet function $S_t(x_i)$ \cite{jet-function-LiHN2002} is expressed as
\begin{eqnarray}
S_t(x_i)=\frac{2^{1+2c}\Gamma(3/2+c)}{\sqrt{\pi}\Gamma(1+c)}[x_i(1-x_i)]^c.
\end{eqnarray}

The running strong coupling  $\alpha_S$\cite{PDG2016} is expressed as
\begin{eqnarray}
\alpha_s(\mu^2)=\frac{\pi}{\beta_0\text{ln}(\mu^2/\Lambda_{QCD}^2)} -\frac{\pi\beta_1\text{ln}(\text{ln}(\mu^2/\Lambda_{QCD}^2))} {\beta_0^3\text{ln}^2(\mu^2/\Lambda_{QCD}^2)}.
\end{eqnarray}



\begin{thebibliography}{99}


\bibitem{Jones2000}
M. K. Jones {\it et al.} (JLab Hall A Coll.), Phys. Rev. Lett. {\bf  84}, 1398 (2000).

\bibitem{Gayou2002}
O. Gayou {\it et al.} (JLab Hall A Coll.),  Phys. Rev. Lett. {\bf 88}, 092301 (2002).


\bibitem{Andivahis1994}
L. Andivahis {\it et al.}, Phys. Rev. D {\bf  50}, 5491 (1994).


\bibitem{Walker1994}
R. C. Walker {\it et al.}, Phys. Rev. D {\bf  49}, 5671 (1994).


\bibitem{Blunden03}P. G. Blunden, W. Melnitchuk, and J. A. Tjon, Phys. Rev. Lett.
\textbf{ 91}, 142304 (2003).

\bibitem{Kondra05}S. Kondratyuk, P.G. Blunden, W. Melnitchuk, and J. A. Tjon, Phys. Rev. Lett.
\textbf{ 95}, 172503 (2005).

\bibitem{Blunden05}P. G. Blunden, W. Melnitchuk, and J. A. Tjon, Phys. Rev. C
\textbf{ 72}, 034612  (2005).

\bibitem{zhouhq2014}
Hong-Yu Chen, Hai-Qing Zhou,
Phys.Rev. C {\bf90}, 045205 (2014).




\bibitem{Chen04}Y. C. Chen, A. Afanasev, S. J. Brodsky, C. E.
Carlson, and M. Vanderhaeghen,  Phys. Rev. Lett. {\bf 93}, 122301.
(2004).

\bibitem{Afana05}A. Afanasev, S. J. Brodsky, C. E. Carlson, Y. C. Chen, and M.
Vanderhaeghen,  Phys. Rev. D {\bf 72}, 013008 (2005).



\bibitem{Chen07}Y. C. Chen, C. W. Kao and S. N. Yang, Phys. Lett. B {\bf 652}, 269 (2007).

\bibitem{BK07}D. Borisyuk and A. Kobushkin, Phys. Rev. C {\bf 76},
022201 (2007).

\bibitem{BK08}D. Borisyuk and A. Kobushkin, Phys. Rev. C {\bf 78},
025208 (2008).

\bibitem{BK06}D. Borisyuk and A. Kobushkin, Phys. Rev. C {\bf  74},
065203 (2006).

\bibitem{BK11} D. Borisyuk and A. Kobushkin, Phys. Rev. C {\bf  83},
057501 (2011).

\bibitem{BK12} D. Borisyuk and A. Kobushkin, Phys. Rev. C {\bf  86},
055204 (2012).

\bibitem{BK14} D. Borisyuk and A. Kobushkin, Phys. Rev. C {\bf  89}, 025204 (2014).

\bibitem{Blunden2017}
 P.G. Blunden, W. Melnitchouk, Phys.Rev. C {bf 95}, 065209 (2017).


\bibitem{BK09}D. Borisyuk and A. Kobushkin, Phys. Rev. C {\bf  79},
034001 (2009).

\bibitem{Kivel09}N. Kivel and M. Vanderhaeghen,  Phys. Rev. Lett. {\bf  103}, 092004 (2009).


\bibitem{TPE-SCEF}
N. Kivel and M. Vanderhaeghen, J. High Energy Phys. {\bf 04}, 029 (2013).


\bibitem{OLYMPUS2017}
B. S. Henderson, Phys. Rev. Lett. {\bf 118}, 092501 (2017)

\bibitem{DianYongChen2008}
D.Y. Chen, H.Q. Zhou, Y.B. Dong, Phys. Rev. C{\bf78},(2008) 045208.

\bibitem{Blunden2010}
P.G. Blunden, W. Melnitchouk,J.A. Tjon, Phys. Rev. C 81 (2010) 018202.


\bibitem{YuBingDong2010}
Yu Bing Dong, S.D. Wanga, Phys. Lett. B 684 (2010) 123.

\bibitem{DianYongChen2013}
Dian-Yong Chen, Yu-Bing Dong,
Phys.Rev. C {\bf 87}, 045209,(2013).



\bibitem{Tomalak2014}
O. Tomalak, M. Vanderhaeghen, Phys.Rev. D {\bf 90}, 013006,  (2014).

\bibitem{Afanasev2016}
O. Koshchii and A. Afanasev, Phys. Rev. D {\bf 94}, 116007 (2016).

\bibitem{zhouhaiqing2017}
Hai-Qing Zhou, Phys.Rev. C {\bf 95}, 025203 (2017).



\bibitem{ZhengTaoWei2003}
Zheng-Tao Wei, Mao-Zhi Yang, Phys. Rev. D {\bf 67}, 094013 (2003).


\bibitem{XingGangWu2004}
Tao Huang {\em et al.}, Phys. Rev. D {\bf 70}, 093013 (2004).


\bibitem{Raha2009-PRD}
Udit Raha {\em et al.}, Phys. Rev. D {\bf 79}, 034015 (2009).


\bibitem{Raha2010-PLB}
J. W. Chen, H. Kohyama, Kazuaki Ohnishi, Udit Rahaa, Yue-Long Shen, Phys. Lett. B {\bf 693},102 (2010).


\bibitem{HaoChungHu2013-PLB}
Hao-Chung Hu, Hsiang-nan Li, Phys. Lett. B {\bf 718},1351 (2013).

\bibitem{ShangChen2015-PLB}
ShanCheng, Zhen-JunXiao, Phys. Lett. B {\bf 749}, 1  (2015).


\bibitem{pQCD}
G.P. Lepage and S.J. Brodsky, Phys. Rev. Lett. {\bf 43}, 545  (1979); G.P. Lepage and S.J. Brodsky, Phys. Rev. D {\bf 22}, 2157  (1980);
J. Botts and G. Sterman, Nucl. Phys. B {\bf 325}, 62  (1989); H-n. Li and G. Sterman, Nucl. Phys. B {\bf 381}, 129 (1992).

\bibitem{Sudkov-factor-Sterman}
J.Botts and G.Sterman, Nucl. Phys. B {\bf 325}, 62 (1989); Hsiang-Nan Li and George Sterman, Nucl. Phys. B {\bf381}, 129  (1992) .

\bibitem{jet-function-LiHN2002}
T. Kurimoto, H. N. Li and A. I. Sanda, Phys. Rev. D{\bf 65}, 014007(2002); H.N. Li, Phys.Rev.D
{\bf 66}, 094010(2002).

\bibitem{Cuba}
T. Hahn, Comput. Phys. Commun. {\bf 168} (2005) 78; T. Hahn, Comput. Phys. Commun. {\bf 207}, 341  (2016).

\bibitem{PDG2016}
Particle Data Group, Chin. Phys. C {\bf40}, 100001  (2016).


\end{thebibliography}
\end{document}